\def\etal{{\frenchspacing\it et al.}}

\def\beq#1{\begin{equation}\label{#1}}
\def\eeq{\end{equation}}
\def\beqa#1{\begin{eqnarray}\label{#1}}
\def\eeqa{\end{eqnarray}}

\def\etal{{\frenchspacing\it et al.}}

\def\fun#1#2{\lower3.6pt\vbox{\baselineskip0pt\lineskip.9pt
        \ialign{$\mathsurround=0pt#1\hfill##\hfil$\crcr#2\crcr\sim\crcr}}}
        
\documentclass[12pt,preprint]{aastex}

\usepackage{graphicx}
\usepackage{epsfig}
\usepackage{epstopdf}

\newcommand{\be}{\begin{equation}}
\newcommand{\ee}{\end{equation}}
\newcommand{\ba}{\begin{eqnarray}}
\newcommand{\ea}{\end{eqnarray}}

\shorttitle{Improving Calibration of SNe~Ia}
\shortauthors{Wang and Hall}

\begin{document}

\title{Improving the Calibration of Type Ia Supernovae\\
Using Late-time Lightcurves}
\author{Yun Wang, and Nicholas Hall}
\affil{Homer L. Dodge Department of Physics \& Astronomy\\ 
		 Univ. of Oklahoma,
                 440 W Brooks St., Norman, OK 73019\\
                 email: wang@nhn.ou.edu, nrhall@ucdavis.edu}

\begin{abstract}

The use of Type Ia supernovae (SNe Ia) as cosmological standard candles
is a key to solving the mystery of dark energy.
Improving the calibration of SNe Ia
increases their power as cosmological standard candles.
We find tentative evidence for a correlation between the
late-time lightcurve slope and the peak luminosity of SNe Ia
in the $B$ band; brighter SNe Ia seem to have shallower 
lightcurve slopes between 100 and 150 days from maximum light.
Using a Markov Chain Monte Carlo (MCMC) analysis in calibrating SNe Ia, 
we are able to simultaneously take into consideration the effect of dust 
extinction, the luminosity and lightcurve width correlation (parametrized 
by $\Delta m_{15}$), and the luminosity and late-time lightcurve slope correlation.
For the available sample of 11 SNe Ia with well-measured late-time 
lightcurves, we find that correcting for the correlation between 
luminosity and late-time lightcurve slope of the SNe Ia
leads to an intrinsic dispersion of 0.12 mag in the Hubble diagram.
Our results have significant implications for future 
supernova surveys aimed to illuminate the nature of dark energy.

\end{abstract}

%\end{document}


\section{Introduction}

One of the most important problems in cosmology today is to
solve the mystery of dark energy, the unknown reason for the
observed cosmic acceleration \citep{Riess98,Perl99}.
The discovery of cosmic acceleration was made using
Type Ia Supernovae (SNe Ia), as SNe Ia can be calibrated
to be good cosmological standard candles 
\citep{Phillips93,Riess95}. Improving the calibration of SNe Ia
increases their power as cosmological standard candles.

A SN~Ia is a thermonuclear explosion that
completely destroys a carbon/oxygen white dwarf  
as it approaches the Chandrasekher limit of 
1.4 $M_{\odot}$.\footnote{The exception recently found by 
\cite{Howell06} is super-luminous
and can be easily separated from the normal SNe~Ia used for cosmology.} 
This is the reason SNe~Ia can be calibrated to be good standard candles.
The first challenge to overcome when using SNe~Ia as cosmological
standard candles is properly incorporating the intrinsic scatter 
in SN~Ia peak luminosity.
The usual calibration of SNe~Ia reduces the intrinsic scatter in SN~Ia peak 
luminosity to about 0.17~mag \citep{Phillips93,Riess95}.
The calibration techniques used so far are based on one observable parameter,
the lightcurve width, which can be parametrized either as $\Delta m_{15}$
(decline in magnitudes in the $B$-band
for a SN~Ia in the first 15 SN Ia restframe days after $B$-band maximum,
see \cite{Phillips93}), or a stretch factor (which linearly scales the time axis, 
see \cite{Goldhaber01}).
The lightcurve width is associated with the amount of $^{56}$Ni
produced in the SN~Ia explosion, which in turn depends on
when the carbon burning makes the transition from turbulent 
deflagration to a supersonic detonation \citep{Wheeler03}.

There are theoretical reasons to expect the existence of
additional physical parameters that characterize the
lightcurve of a SN~Ia.
For example, \cite{Milne01} modeled the late light
curves of SNe~Ia, and found that the nonlocal and time-dependent
energy deposition due to the transport of Comptonized electrons can
produce a 0.10 to 0.18~mag correction to the late-time brightness of
SNe~Ia, depending on their position in the $\Delta m_{15}$ 
sequence.

In this paper we show that late-time lightcurves of SNe Ia
can be used to tighten the calibration of SNe~Ia, and reduce their
effective (post-calibration) intrinsic scatter
in the Hubble diagram. We describe our method in Section 2, and
present our results in Section 3. We summarize and conclude in
Section 4.

\section{Method}

\subsection{Defining the sample}

We use the lightcurve fits obtained using the 
Joint SuperStretch code for SN Ia lightcurves \citep{Lifan06}
\footnote{The B and V lightcurves were fit using no secondary bump.}
to derive the lightcurve parameters (epoch and 
apparent magnitudes at maximum light and $\Delta m_{15}$). 
This is necessary for 
consistency in making color and $\Delta m_{15}$ corrections.

We select a sample of ``normal'' SNe Ia (from a total of 135 available) 
that satisfy the following requirements:

\noindent
(1) $B_{max}-V_{max}\leq 0.5 $;\\
\noindent
\noindent
(2) $0.015 < z < .1$;\\
\noindent
(3) B lightcurve data extending beyond 100 days from maximum light.

The requirement of $B_{max}-V_{max}\leq 0.5 $ selects
the ``Branch normal'' SNe Ia \citep{Vaughan95,branch98}, which
is standard in studying SNe Ia as cosmological standard candles.
The requirement of $0.015 < z < .1$ selects SNe Ia
which are nearby yet distant enough so that they
are not affected by peculiar velocities due to cosmic large
scale structure that modifies the Hubble diagram of
nearby standard candles \citep{Wang98}, which
has the largest impact on the nearest SNe~Ia (see, e.g., 
\cite{Zehavi98,Cooray06,Hui06,Conley07}).

This yields a sample of 13 SNe Ia. 
Of these, we exclude
two more SNe: SN1999cc and SN1993h.
%SN2001ay has an average luminosity, but an abnormally small decline rate 
%of $\Delta m_{15}=0.54$.
SN1999cc has only one data point in the B lightcurve between 40 and 100 days 
from maximum light, insufficient to determine a slope 
independent of the lightcurve model.
For SN1993h, the late-time lightcurve
slope appears to depend on the lightcurve model, due to the
appearance of a small secondary peak in the lightcurve data at $\sim$ 
80 days from maximum light.
For the final set of 11 SNe Ia, the B lightcurves are very close to 
a straight line at $\ga$ 60 days from maximum light in the 
SN Ia restframe, and there are a minimum of two data points 
at $t\ga 60$ days from maximum light.

Our final sample consists of 11 SNe Ia that have lightcurve
data at $\ga$ 100 days from maximum light in the 
SN Ia restframe, and late-time lightcurve
slopes insensitive to the lightcurve model.
Table 1 lists the SNe Ia in our sample.
We include all the data that are required to allow others
to reproduce our results. Note that $\Delta b_{lt}$
and $N_{lt}$ are {\it not} required to reproduce our
results [see Eqs.(\ref{eq:fit})-(\ref{eq:chi2})], but
are given for reference.
\begin{table*}[htb]
\caption{Nearby SNe Ia with well defined late-time lightcurves. }
\begin{center}
\begin{tabular}{llllllllll}
\hline
SN name & $z$ & $B_{max}$ & $\Delta B_{max}$ &$V_{max}$ & $\Delta m_{15}$ & $b_{lt}$ 
& $\Delta b_{lt}$& $N_{lt}$ & reference \\
\hline
SN1990O  & 0.031 & 16.607 & 0.035 & 16.533 & 0.961 & 0.0160 & 0.0008 & 4 &H96b\\ 
SN1990T  & 0.040 & 17.454 & 0.086 & 17.537 & 1.298 & 0.0152
& 0.0028 & 6 &H96b\\ 
SN1990Y  & 0.039 & 17.674 & 0.306 & 17.494 & 1.217 & 0.0153 & 0.0025 & 7 &H96b \\     
SN1992bc & 0.020 & 15.207 & 0.009 & 15.207 & 0.960 & 0.0147 & 0.0002 & 10 &H96b \\        
SN1993ag & 0.050 & 18.298 & 0.026 & 18.081 & 1.382 & 0.0142 & 0.0030 & 3 &H96b\\  
SN1997dg & 0.033 & 17.143 & 0.033 & 17.117 & 1.281 & 0.0177 & -- & 2 &J06\\ 
SN1998ec & 0.020 & 16.670 & 0.053 & 16.406 & 1.074 & 0.0164 & 0.0033 & 4 &J06\\ 
SN1998V  & 0.017 & 15.912 & 0.029 & 15.738 & 1.150 & 0.0163 & 0.0012 & 5 &J06\\ 
SN1999ef & 0.038 & 17.524 & 0.117 & 17.418 & 1.055 & 0.0160 & 0.0049 & 3 &J06\\         
SN2000dk & 0.016 & 15.637 & 0.021 & 15.543 & 1.457 & 0.0136
& 0.0029 & 4 &J06\\        
SN2000fa & 0.022 & 16.026 & 0.051 & 16.055 & 1.140 & 0.0164 & 0.0028 & 4 &J06 \\   
 \hline		
\end{tabular}
\tablecomments{Note that H96b refers to \cite{Hamuy96b}, and J06 refers to \cite{jha06}.
$N_{lt}$ refers to the number of data points in the lightcurve at $t\ga 60$ days
from maximum light.}
\end{center}
\end{table*}

\subsection{Fitting the Absolute Magnitudes}

To convert observed apparent magnitudes to absolute magnitudes 
we assume a flat universe, with
$\Omega_m=0.3$, $\Omega_{\Lambda}=0.7$,
and $H_0 = 70\,$km$\,$s$^{-1}\,$Mpc$^{-1}$.
The B absolute magnitudes are 
\be
M_B^{obs} \equiv B_{max} - \mu,
\ee
where $B_{max}$ denotes the apparent magnitude at peak brightness
in the B band, and the distance modulus $\mu$ is 
\be
\mu(z) = m-M=5 \log \left[d_L(z)/10{\,{\mbox{pc}}}\right].
\ee
For $z<0.1$, the luminosity distance $d_L(z)$ is well approximated by
\be
d_L(z)= \frac{cz}{H_0}\,\left[1+ \frac{1}{2}(1-q_0)\,z \right],
\label{eq:d_L}
\ee
with $q_0=\Omega_m/2-\Omega_\Lambda$. 
Since we are only considering very low redshift SNe Ia,
our results are insensitive to the assumed cosmological model.

Without any corrections, the dispersion of the SNe Ia in our
sample in the Hubble diagram is 0.35 mag (see Fig.5(a)).
This is due to extinction by dust, as well as the
intrinsic dispersion in the luminosities of SNe Ia due to
the physical diversity of these objects \citep{Woo07}.
Our goal is to model these effects phenomenologically 
using observational data, in order to calibrate SNe Ia
into good standard candles.

According to \cite{Card89}, due to dust extinction,
all true B magnitudes differ from
the observed B magnitudes by $\delta B\equiv B-B^{obs}
=R_B\,\delta(B-V)$, where $\delta(B-V)\equiv (B-V)-(B-V)^{obs}$, and
$R_B$ is approximately constant.
Following \cite{vandenBergh95}, we use ${\cal R}$ to denote the color
correction coefficient parametrizing the combined effects of
dust extinction and intrinsic SN color variation.
Assuming that $B_{max}-V_{max}=0$ for unreddened SNe Ia \citep{Phillips99},
the dust correction to $M_B^{obs}$ is ${\cal R} E$, with
$E=B_{max}-V_{max}$.

There is a well known correlation between SN Ia peak luminosity
and $\Delta m_{15}$, the decline in brightness (in units of mag)
15 days (in the SN restframe) after peak brightness \citep{Phillips93}.
Following \cite{Lifan06}, we parametrize the $\Delta m_{15}$
dependence of $M_B^{obs}$ with a piecewise linear function,
${\cal D}(\Delta m_{15})=\alpha\,\left(\Delta m_{15}-1.1\right)+
\alpha\prime\, \left|\Delta m_{15}-1.1\right|$.

We use the Markov Chain Monte Carlo (MCMC) technique (see
\cite{neil} for a review),
illustrated for example in \cite{Lewis02}, in the likelihood analysis.
The MCMC method scales approximately linearly in computation
time with the number of parameters. 
The method samples from the full posterior distribution of the
parameters, and from these samples the marginalized posterior distributions
of the parameters can be estimated. 
In the MCMC analysis, we fit $M_B^{obs}$ to 
\be
M_B^{fit}=M+\Delta M^{cor}=M + {\cal R} E + \alpha\,\left(\Delta m_{15}-1.1\right)+
\alpha\prime\, \left|\Delta m_{15}-1.1\right| + \beta b_{lt},
\label{eq:fit}
\ee
where $b_{lt}$ (in units of mag/day) is the late-time 
lightcurve slope (in the SN restframe). In our MCMC code, we use
\be
\chi^2= \sum_{i=1}^{N} \left(\frac{M_B^{obs}(z_i)-M_B^{fit}(z_i)}{\Delta B_{max}(z_i)}
\right)^2.
\label{eq:chi2}
\ee
The MCMC algorithm allows us to explore the dependence
of $M_B^{obs}$ on all the parameters simultaneously.

The Hubble diagram dispersion is a conventional measure of 
the intrinsic scatter of SNe Ia.  It is
given by the standard deviation of the 
apparent magnitudes at maximum light of the SNe Ia, $B_{max}$, 
from the theoretical predictions given by
\be
m_0(z) \equiv \mu(z) + M,
\ee
where $M$ is a constant that denotes 
the absolute magnitude of SNe Ia.  It can only be determined 
through direct measurements of the distances to the SNe Ia
using Cepheid variable stars. 
Since $M$ has no impact on the relative calibration
of SNe Ia that we study in this paper, we allow $M$ to vary
in order to obtain the smallest scatter in the Hubble diagram,
independent of the $M$ values obtained in the MCMC likelihood analysis.
This is the same approach adopted by \cite{Lifan06}
using a different likelihood analysis technique.

\section{Results}

We have used the restframe fitted lightcurves to obtain slopes of the 
lightcurves at late-time, chosen to be between 100 and 150 
days from maximum light.  
Since the lightcurves are very close to straight lines at $\ga 60\,$days
from maximum light, the measured slopes are robust and not sensitive
to the lightcurve model. Table 1 lists the late-time lightcurve slopes
$b_{lt}$ for the SNe Ia in our sample.\footnote{The uncertainties in
$b_{lt}$ are estimated from fitting the data points in the lightcurve
at $t\ga 60\,$days from maximum light to a straight line. Note that
$\Delta b_{lt}$ values are {\it not} used in our analysis,
see Eqs.(\ref{eq:fit})-(\ref{eq:chi2}).}

Our data consist of ($B_{max}$, $\Delta B_{max}$, $E$, $\Delta m_{15}$, $b_{lt}$)
for each of the SNe Ia in our sample.
The parameters that we vary in our MCMC likelyhood analysis
are ($M$, ${\cal R}$, $\alpha$, $\alpha\prime$, $\beta$).
\cite{Lifan06} found that ${\cal R}=2.59\pm0.24$, $\alpha=0.78\pm 0.09$,
and $\alpha\prime=-0.54\pm 0.19$. To avoid obtaining unphysical
values of $\alpha$ and $\alpha\prime$ due to the small size
of our sample, we impose a prior of $\alpha\prime \ge -2$ in our analysis.

We obtain millions of MCMC samples, which are then appropriately
thinned for statistical independence.
We do not present results on the ``nuisance'' parameter
$M$; it is marginalized over in obtaining the probability
distributions that we present.
Fig.1 shows the marginalized probability density functions (pdf)
of ${\cal R}$, $\alpha$, $\alpha\prime$, and $\beta$.
Fig.2 shows the corresponding 2-D joint confidence contours
(68\% and 95\%) for these parameters.
For comparison, Fig.3 and Fig.4 show the pdf and joint confidence
contours for ${\cal R}$, $\alpha$, and $\alpha\prime$, with $\beta=0$.
Fig.5 and Fig.6 show the pdf and joint confidence
contours for $\alpha$, $\alpha\prime$, and $\beta$, for
fixing ${\cal R}=4.1$ (standard Milky-Way dust).
In Fig.1, Fig.3, and Fig.5, the dotted lines are the likelihood functions.
Agreement between the pdf and the likelihood function indicates
that the MCMC chains have converged and hence provide reliable
pdf's.
Table 2 lists the best fit, mean, and standard deviations for
${\cal R}$, $\alpha$, $\alpha\prime$, $\beta$.

\begin{figure}
\psfig{file=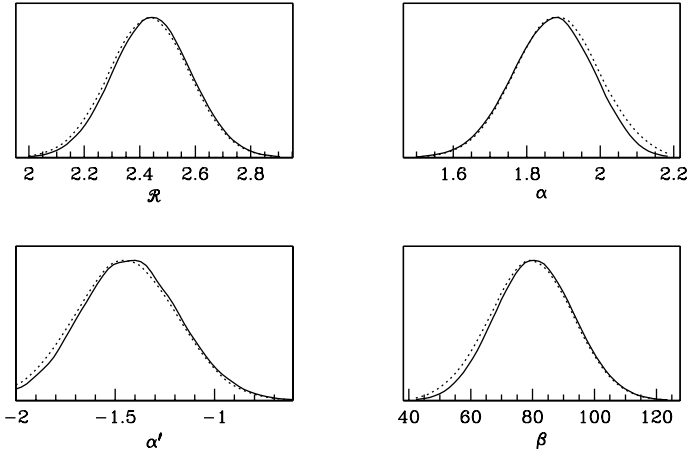,width=4in,height=4in}
\vskip-3cm
\caption{Marginalized probability density functions (pdf)
of ${\cal R}$, $\alpha$, $\alpha\prime$, and $\beta$.
}
\label{fig1}
\end{figure}

\begin{figure}
\psfig{file=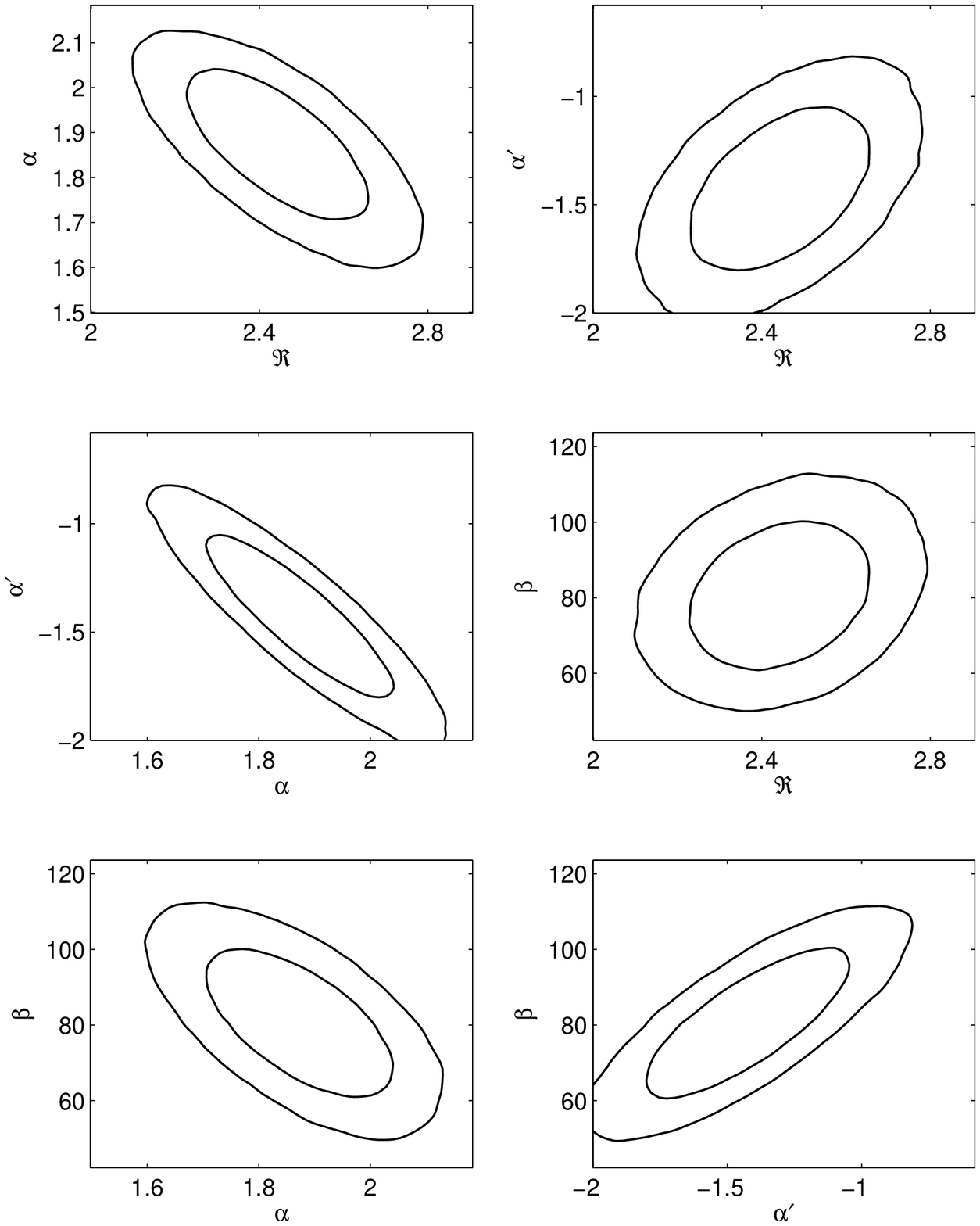,width=4in,height=4in}
%\includegraphics{f2.eps}
%\vskip-3cm
\caption{2-D joint confidence level contours
(68\% and 95\%) corresponding to Fig.1.
}
\label{fig2}
\end{figure}

\begin{figure}
\psfig{file=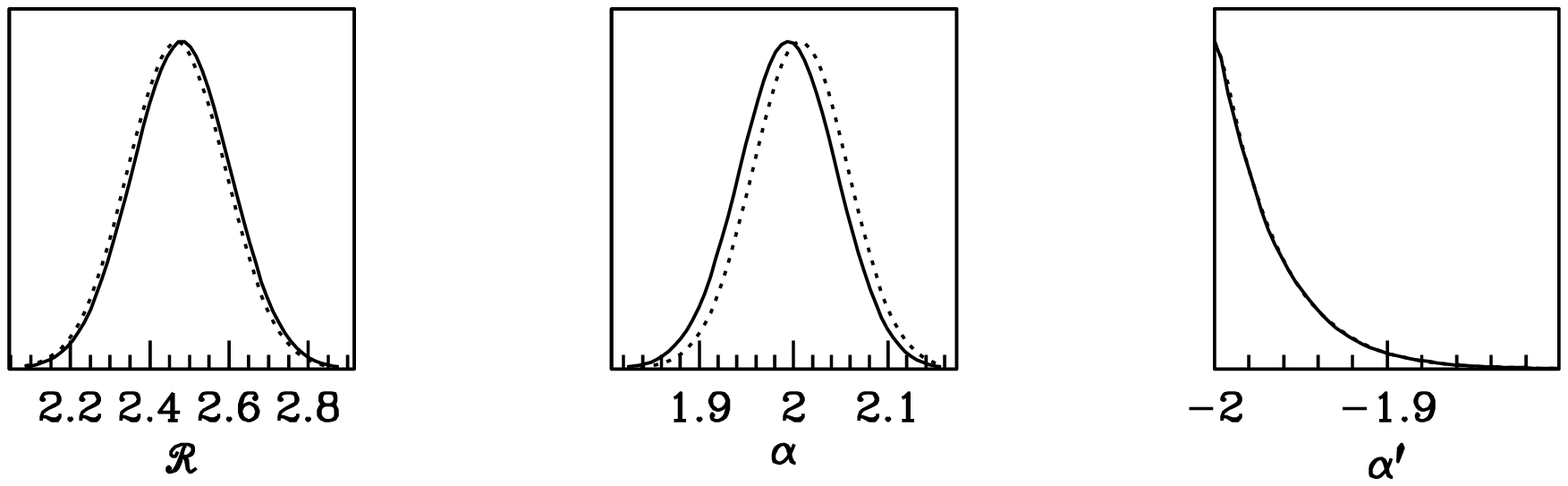,width=5in,height=4in}
\vskip-5cm
\caption{Marginalized probability density functions (pdf)
of ${\cal R}$, $\alpha$, and $\alpha\prime$, for $\beta=0$.
}
\label{fig3}
\end{figure}

\begin{figure}
\psfig{file=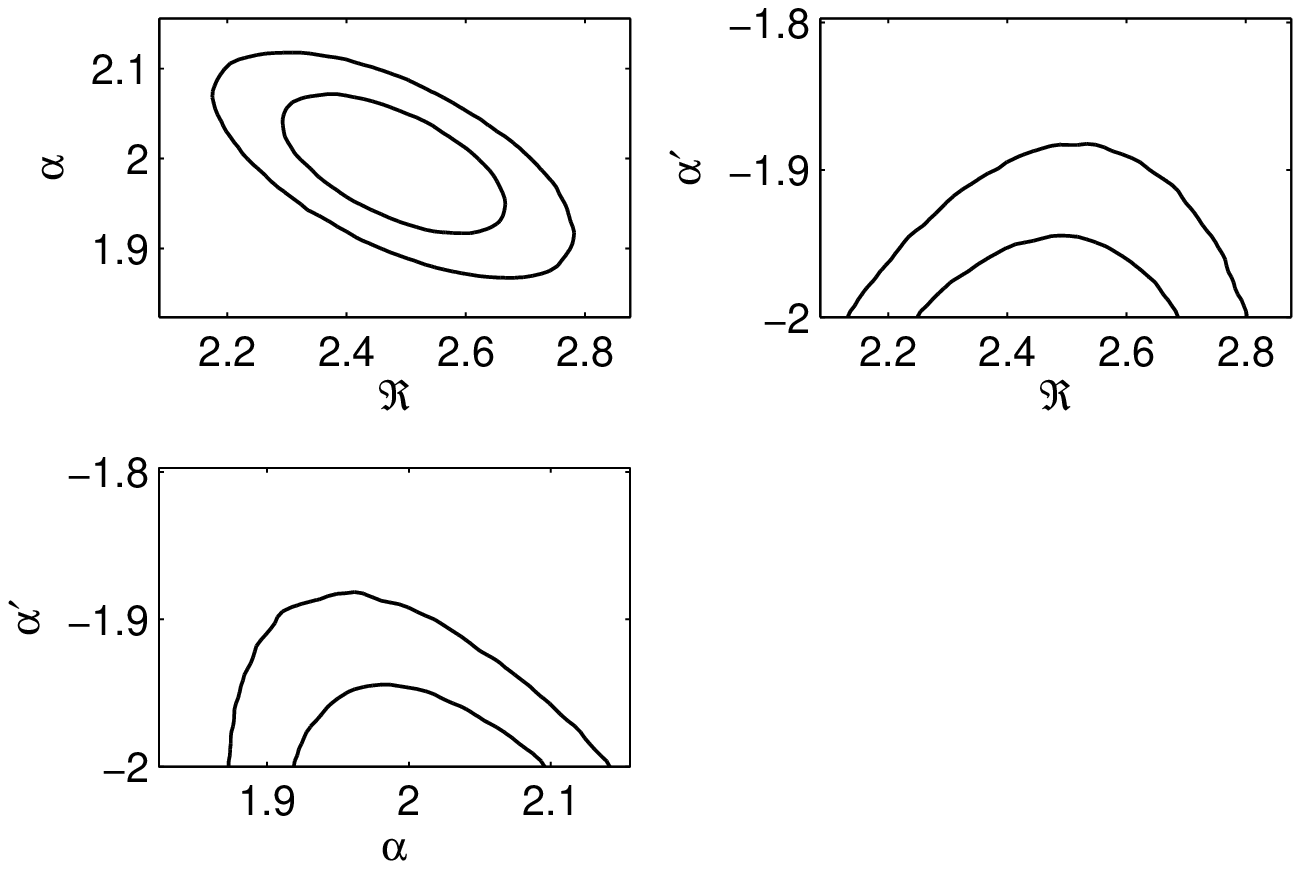,width=4in,height=4in}
\caption{2-D joint confidence level contours
(68\% and 95\%) corresponding to Fig.3.
}
\label{fig4}
\end{figure}

\begin{figure}
\psfig{file=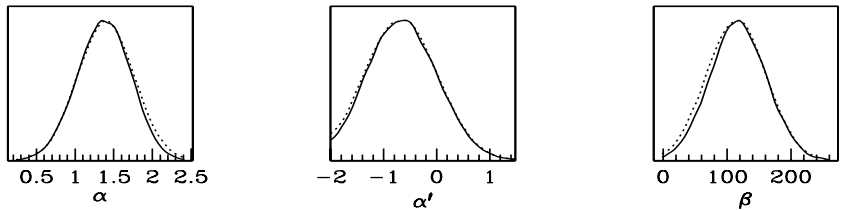,width=5in,height=4in}
\vskip-5cm
\caption{Marginalized probability density functions (pdf)
of $\alpha$, and $\alpha\prime$, and $\beta$,
for fixing ${\cal R}=4.1$ (standard Milky-Way dust).
}
\label{fig5}
\end{figure}

\begin{figure}
\psfig{file=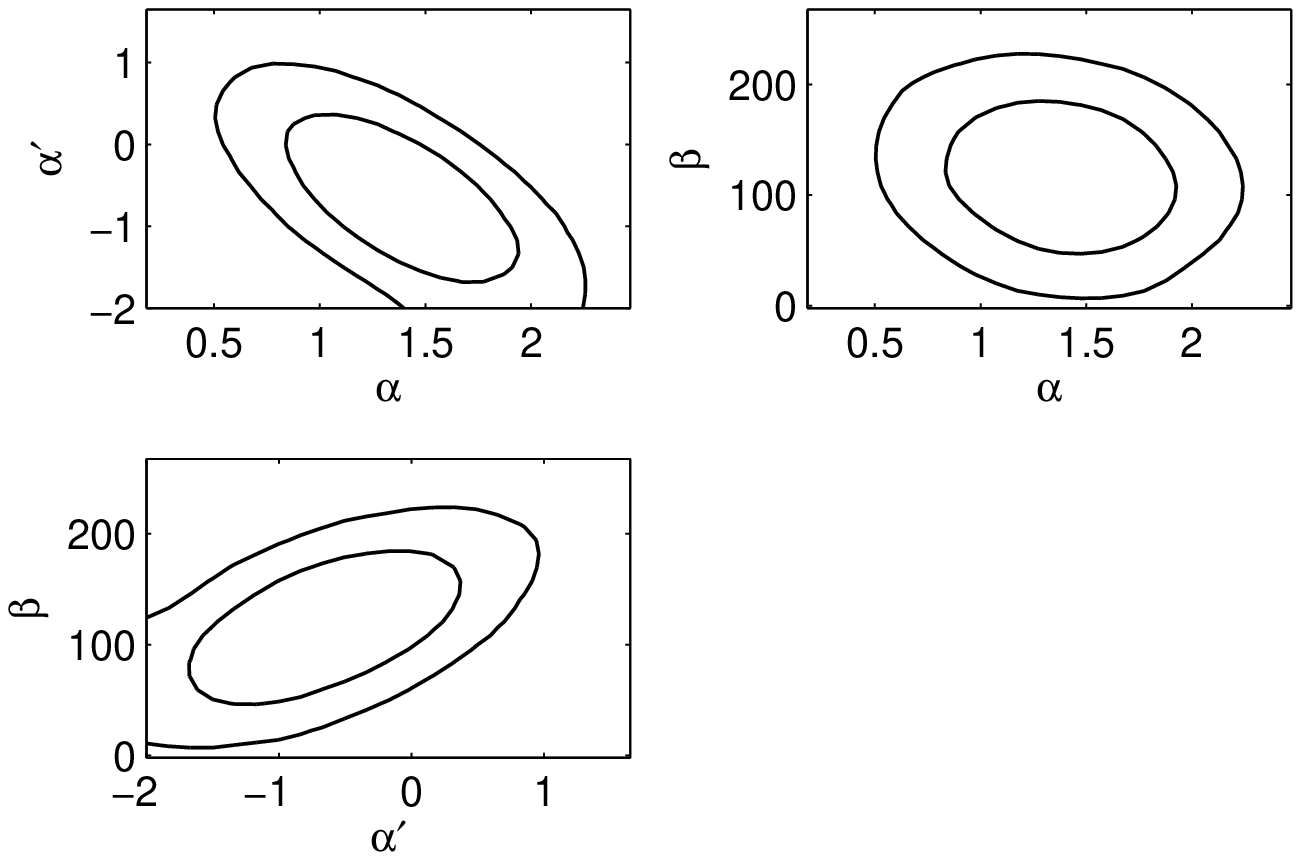,width=4in,height=4in}
\caption{2-D joint confidence level contours
(68\% and 95\%) corresponding to Fig.5.
}
\label{fig6}
\end{figure}

\begin{table*}[htb]
\caption{Best fit, mean, and standard deviations for
${\cal R}$, $\alpha$, $\alpha\prime$, $\beta$.}
\begin{center}
\begin{tabular}{lllll}
\hline
parameter & best fit  & mean & rms dev \\
\hline
\hline
&including $b_{lt}$: && \\
\hline
 ${\cal R}$  &  2.407 & 2.445 & 0.141 \\
 $\alpha$  & 1.890 &1.870 & 0.109\\
 $\alpha\prime$  & $-$1.453 &$-$1.418  &0.244\\
 $\beta$  & 77.490 &80.925 & 12.779\\
\hline
\hline
& not including $b_{lt}$ & ($\beta=0$):&\\
\hline
 ${\cal R}$  & 2.454 & 2.480 & 0.123\\
 $\alpha$  &  2.007 &   1.994 & 0.051\\
 $\alpha\prime$ & $-$2.000 & $-$1.969 & 0.030\\
 \hline	
 \hline
& forcing  ${\cal R}=4.1$: & &\\
\hline
 $\alpha$  &1.372  & 1.371 & 0.345 \\
 $\alpha\prime$  & $-$0.583  & $-$0.621  & 0.633 \\
 $\beta$ & 112.364  & 117.889 & 43.939 \\
 \hline		
\end{tabular}
%\tablecomments{}
\end{center}
\end{table*}

It is reassuring that we obtain the same value of ${\cal R}$ whether
or not we include late-time lightcurve slope $b_{lt}$ in our analysis.
Note that the prior of $\alpha\prime \geq -2$ has little impact if
we include $b_{lt}$ in our analysis (see Figs.1-2), but it
has a very large impact if we only fit for ($M$, ${\cal R}$, $\alpha$, $\alpha\prime$).
This indicates that the set of parameters ($M$, ${\cal R}$, $\alpha$, $\alpha\prime$)
does not properly describe the data, given reasonable physical
priors on $\alpha\prime$.

Fig.7 shows the Hubble diagrams for three cases,
with the solid line indicating $m_0(z)=\mu(z)+M$ in each case,
where $M$ is obtained by minimizing the rms of $|B_{max}(z)-m_0(z)|$
(the Hubble diagram dispersion).
The three cases are:
(a) No corrections. 
(b) $B_{max}$ corrected by subtracting 
$\Delta M^{cor}={\cal R} E + \alpha\,\left(\Delta m_{15}-1.1\right)+
\alpha\prime\, \left|\Delta m_{15}-1.1\right|$, with
${\cal R}=2.45$, $\alpha=2.01$, and $\alpha\prime=-2.00$.
(c) $B_{max}$ corrected by subtracting 
$\Delta M^{cor}={\cal R} E + \alpha\,\left(\Delta m_{15}-1.1\right)+
\alpha\prime\, \left|\Delta m_{15}-1.1\right|+ \beta b_{lt}$, with
${\cal R}=2.41$, $\alpha=1.89$, $\alpha\prime=-1.45$,
and $\beta=77.49$ (mag/day)$^{-1}$.
The parameter values in (b) and (c)
correspond to the best fit values from the MCMC analysis
(listed in Table 2).
We find that including the correlation between luminosity
and late-time lightcurve slope in correcting $M_B^{obs}$ reduces the 
Hubble diagram dispersion from 0.14 mag to 0.12 mag.
For comparison, fixing ${\cal R}=4.1$ (standard Milky-Way dust)
gives a Hubble diagram dispersion of 0.22 mag. This indicates that 
the assumption of standard Milky-Way dust is not supported
by the data.

\begin{figure}
\psfig{file=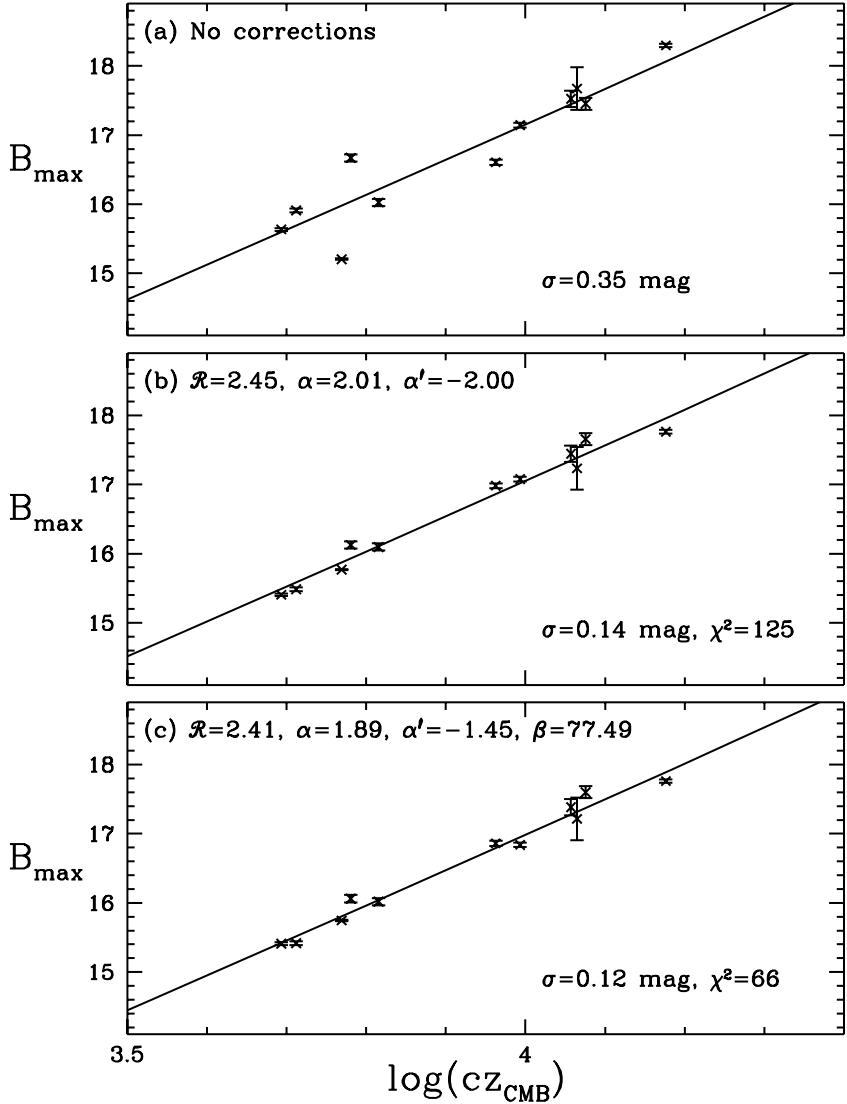,width=6in,height=6in}
\caption{Hubble diagrams for three cases, with the solid line indicating 
$m_0(z)=\mu(z)+M$ in each case, where $M$ is obtained by minimizing the 
rms of $|B_{max}(z)-m_0(z)|$ (the Hubble diagram dispersion).
(a) No corrections. 
(b) $B_{max}$ corrected by subtracting 
$\Delta M^{cor}={\cal R} E + \alpha\,\left(\Delta m_{15}-1.1\right)+
\alpha\prime\, \left|\Delta m_{15}-1.1\right|$, with
${\cal R}=2.45$, $\alpha=2.01$, and $\alpha\prime=-2.00$.
(c) $B_{max}$ corrected by subtracting 
$\Delta M^{cor}={\cal R} E + \alpha\,\left(\Delta m_{15}-1.1\right)+
\alpha\prime\, \left|\Delta m_{15}-1.1\right|+ \beta b_{lt}$, with
${\cal R}=2.41$, $\alpha=1.89$, $\alpha\prime=-1.45$,
and $\beta=77.49$ (mag/day)$^{-1}$.
For comparison, fixing ${\cal R}=4.1$ (standard Milky-Way dust)
gives a Hubble diagram dispersion of $\sigma=0.22$ mag, much larger
than the Hubble diagram dispersions in panels (b) and (c).
}
\label{fig7}
\end{figure}

To study the effect of the measurement uncertainty
in $B_{max}$ for our sample, we have repeated our MCMC analysis
setting $\Delta B_{max}$ to a constant. 
This gives similar results as above.
We find that including the correlation between luminosity
and late-time lightcurve slope in correcting $M_B^{obs}$ reduces the 
Hubble diagram dispersion from 0.13 mag to 0.11 mag.
This is similar to what we find when using the
actual $\Delta B_{max}$ from Table 1.

Fig.8 shows the absolute magnitude corrected for color and
$\Delta m_{15}$ only [case (b) of Fig.7], versus $b_{lt}$,
with its coefficient $\beta$ given by case (c) of Fig.7.
This shows that the evidence for the correlation between
luminosity and late-time lightcurve slope is tentative
due to the small size of the sample. A much larger sample
of SNe Ia with well measured late-time lightcurves
is needed to firmly establish this correlation.

\begin{figure}
\psfig{file=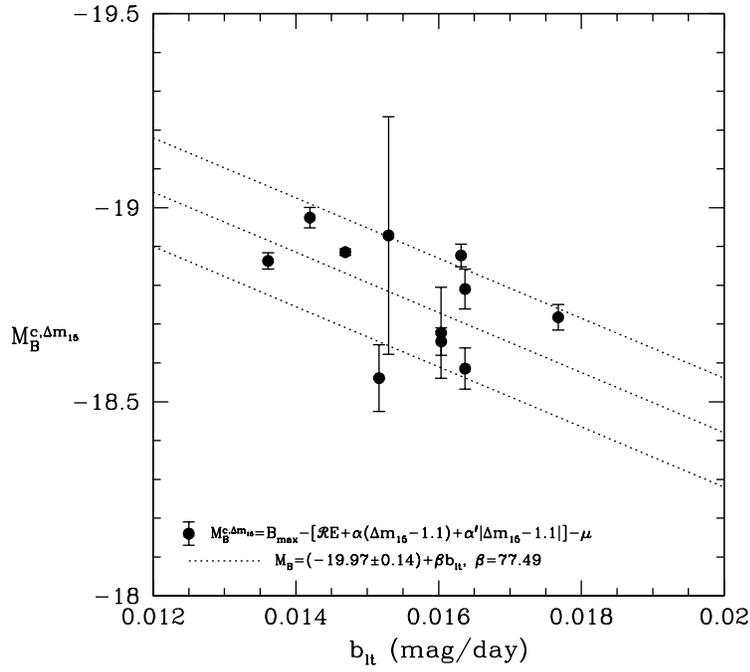,width=4in,height=4in}
\caption{The absolute magnitude corrected for color and
$\Delta m_{15}$ only [case (b) of Fig.7], versus the
late-time lightcurve slope $b_{lt}$,
with its coefficient $\beta$ given by case (c) of Fig.7.
}
\label{fig8}
\end{figure}

\section{Discussion and Summary}
\label{sec:discussion}

Improving the calibration of SNe Ia increases the power of SNe Ia
as cosmological standard candles. 
We have shown the tentative evidence for a correlation between 
SN Ia peak luminosity and late-time lightcurve slope in the $B$ band; 
brighter SNe Ia have shallower lightcurve slopes between 100 and 150 days 
from maximum light (see Fig.8). 
Using this correlation, we are able to reduce the intrinsic scatter
of SNe Ia (measured by the Hubble diagram residue) to 0.12 mag
(see Fig.7).

We use the MCMC algorithm in our likelihood analysis; this
allows us to explore all the parameters simultaneously and efficiently 
in calibrating the SN Ia peak luminosity for our sample of SNe Ia.
We only used $\Delta B_{max}$ in
calculating $\chi^2$, since including the errors in
color, $\Delta m_{15}$, and late-time lightcurve slope
introduces spurious pdfs in the MCMC analysis.
Our results are not sensitive to the uncertainty
in $B_{max}$ (setting $\Delta B_{max}$ to a constant gives
similar results to using the $\Delta B_{max}$ from Table 1).
The parameter set is ($M$, ${\cal R}$, $\alpha$, $\alpha\prime$, $\beta$),
which contains the nuisance parameter and correction coefficients for 
color, $\Delta m_{15}$, and late-time lightcurve slope respectively.
The pdf's we have found for (${\cal R}$, $\alpha$, $\alpha\prime$, $\beta$)
are very close to being Gaussian (see Fig.1 and Fig.3), except for the pdf's
for $\alpha\prime$ (which are cutoff by the prior of $\alpha\prime\geq -2$).
Late-time lightcurve slope $b_{lt}$ appears uncorrelated with color
(see Fig.2), which supports $b_{lt}$ being an intrinsic observable
that correlates with the peak luminosity of SNe Ia. 

The apparent correlation between peak luminosity and late-time lightcurve
slope in the $B$ band that we have found is consistent with the
results by \cite{Capp97}.  They studied the late-time lightcurves 
of five SNe Ia in the $V$ band with the focus of modeling lightcurves
at very late times (later than 150-200 days from maximum light), 
and found a similar trend in the peak luminosity and late-time 
lightcurve slope (which they parametrized with the difference
in $V$ magnitude from maximum to 300 days, see Table 1 of \cite{Capp97}).

The coefficients for correcting for the correlation between
decline-rate and peak luminosity that we have found are very different
from previous work (see Table 2 and
\cite{Phillips93,Hamuy96a,Phillips99,Lifan06}).
This is likely partly a consequence of the particular sample of 
11 SNe Ia we have used for our analysis. This is not surprising,
since previous work by various groups also found very different
$\Delta m_{15}$ corrections \citep{Phillips99,Lifan06},
indicating that this correction is highly sensitive to
sample selection.

Note that the color correction coefficient we found
was not sensitive to whether we correct for late-time
lightcurve slope, and is consistent with 
the color correction coefficient of ${\cal R}=2.59\pm 0.24$ 
found by \cite{Lifan06}.
This is very different from a coefficient of ${\cal R}=4.1$
that is typical of Milky Way dust.
We found that fixing ${\cal R}=4.1$ gives a Hubble diagram residue
almost twice as large as the Hubble diagram residue found
when allowing ${\cal R}$ to vary (see Fig.7 and its caption). This 
strengthens the evidence that the color correction coefficient ${\cal R}$
of SN data cannot be explained by extinction by
typical Milky-Way dust.
The deviation of ${\cal R}$ from ${\cal R}=4.1$ could be due to
the mixing of intrinsic SN Ia color variation with
dust extinction \citep{Conley07}, or variations
in the properties of dust. The definitive resolution 
of this issue will likely require NIR observations of 
a large number of SNe Ia \citep{KK04,Phillips06}.

Our results show the benefit of extending the observations of 
SN Ia lightcurves beyond 100 days in the SN Ia restframe, with
several data points between 50 to 150 days to enable a robust
measurement of the late-time lightcurve slope.
Our sample only contains 11 SNe Ia, limited by currently available
data. 
The expansion in the scope of SN observations to obtain
late-time lightcurves is straightforward
for surveys which cover the same areas in the sky in regular
intervals \citep{Wang00,Phillips06}.

Our results have significant implications for the survey strategy
of future SN surveys, such as 
those planned for the Advanced Liquid-mirror Probe for Astrophysics, 
Cosmology and Asteroids (ALPACA)\footnote{http://www.astro.ubc.ca/LMT/alpaca/};
Pan-STARRS\footnote{http://pan-starrs.ifa.hawaii.edu/}; 
the Dark Energy Survey (DES)\footnote{http://www.darkenergysurvey.org/};
the Large Synoptic Survey Telescope (LSST)\footnote{http://www.lsst.org/},
and the Joint Dark Energy Mission (JDEM).

\bigskip

{\bf Acknowledgments}
We are grateful to Lifan Wang for providing the lightcurve fits
to the SN data we used, Alex Conley for sending us his compilation
of nearby SN data, and David Branch for helpful discussions
and comments on drafts of the manuscript.
This work was supported in part by the NSF CAREER grant AST-0094335
and AST 05-06028.

\end{document}